\documentclass[aps,showpacs,floatfix,a4paper,pre,twocolumn,groupedaddress]{revtex4}

\usepackage[english]{babel}
\usepackage[reqno]{amsmath}
\usepackage{amssymb,mathrsfs}
\usepackage[latin1]{inputenc}

\usepackage{natbib}

\usepackage{multirow}

\usepackage{graphicx}
\usepackage{color}
\graphicspath{{Figs/}}

\usepackage{bm,bbm}
\usepackage{psfrag}
\usepackage[font=small,labelfont=bf]{caption}

\newcommand{\bom}{ \mbox{ \hspace{-1mm}\boldmath $\omega$ \hspace{-1mm}}}

\newcommand{\ba}{\begin{eqnarray}}
\newcommand{\ea}{\end{eqnarray}}

\def\pt{{\partial}}

\def\eps{\varepsilon}
\def\ev{{\bf e}}

\def\cb{\bar{c}}

\def\q1{{q^{(1)}}}

\def\Qv{{\bf Q}}

\def\d{{\rm d}}

\def\elle{{\cal{L}}}

\def\grads{\nabla\hspace{-1mm}_s}

\def\esse{{\cal{S}}}

\def\bom{\vect{\omega}}

\def\ab{\bar a}
\def\abc{\bar a_{\rm cr}}
\newcommand{\vect}[1]{\boldsymbol{\mathbf{#1}}}

\def\Tc{T_{\rm cr}}
\def\TNI{T_{\rm NI}}

\begin{document}

\title{Cooling a spherical nematic shell}
\author{Gaetano Napoli }
\affiliation{Dipartimento di Matematica e Fisica "E. De Giorgi", Universit\`a del Salento, Lecce, Italy}
\author{Luigi Vergori} 
\affiliation{Dipartimento di Ingegneria, Universit\`a degli Studi di Perugia, Perugia, Italy}

\begin{abstract}
Within the framework of Landau-de Gennes theory for nematic liquid crystals, we study the temperature-induced isotropic-nematic phase transition on a spherical shell. Below a critical temperature, a thin layer of nematic coating a microscopic spherical particle exhibits non-uniform textures due to the geometrical frustration. We find the exact value of critical threshold for the temperature and determine exactly the nematic textures at the transition by means of a weakly nonlinear analysis.  The critical temperature is affected by the extrinsic curvature of the sphere, and the nematic alignment  is consistent with the Poincar\'e-Hopf index theorem and experimental observations.  The stability analysis of the bifurcate textures at the isotropic-nematic transition highlight that only the tetrahedral  configuration  is stable.
\end{abstract}

 \maketitle
 Many physical systems exhibit intriguing patterns and textures whose understanding and control is a central goal across many areas of science and engineering. One category of pattern-forming is provided by nematic shells, which consist in nematic liquid crystals (LCs) confined on a self-closing spherical shell. The constraint imposed by such confinement combined with degenerate anchoring of the nematic at the interfaces inevitably results in the presence of topological defects,  i.e. regions where the orientational order of the liquid crystal is disrupted. The seminal idea of Nelson  \cite{Nelson:2002} of mimicking atomic properties at microscopic scale by functionalizing defects of nematic shells has led to a vivid research activity in the last two decades, both theoretical \cite{Vitelli:2006, Biscari:2006, Bates:2008, Shin:2008, Bates:2010, Kralj:2011, nave:2012, naveprl:2012, Segatti:2014, Jesenek:2015, Mesarec:2017} and experimental \cite{Fernandez-Nieves:2007, Lopez-Leon:2011, Liang:2011, Lopez-Leon:2011prl}.  Moreover, recents advances in microfluidics techniques \cite{Lopez-Leon:2011, Liang:2011, Durey:2020, Noh:2020} have made it possible to prepare long-term stable liquid crystalline (LC) shells which, in turn, has  enabled the application of LC shells in the design of innovative devices in the fields of photonics \cite{Geng:2016, Ji-Hwan:2017}, biosensing \cite{Sharma:2019} and micromechanics \cite{Jampani:2018}.

%

Nematic LCs are fluids with orientational order and no long-range positional order.  The Landau-de Gennes theory for nematic LCs  (the most widely used model to describe the thermomechanical response of LCs) is based on the orientational probability distribution of molecules and uses a symmetric traceless second order tensor  $\Qv$ to characterize this distribution. For  nematic LCs occupying a thin shell it has been shown  that, upon appropriate chemical treatments of the shell interfaces,   a fully tangential, a fully normal or a hybrid  alignment can be achieved in the bulk \cite{Sharma:2018}. Within the hypothesis of fully tangential alignment and constant thickness, a surface free energy for nematic shells has been derived from the Lanadu-de Gennes model by means of a perturbation technique  \cite{nave:2012}. In what follows we shall refer to this surface free energy as the surface Landau-de Gennes potential.
Within this setting $\Qv$ degenerates  consistently with the tangential distributions of the average directions of the molecules. These degenerate states are described by the tangential order tensor $\Qv_s$ which, with respect to a local basis on the shell surface, is represented by a $2 \times 2$ traceless symmetric matrix \cite{Biscari:2006}. 

We shall henceforth  consider a spherical surface, and let   $\phi \in [0,2 \pi)$ and $\theta \in [0, \pi]$ be the longitude and the colatitude, respectively. With respect to the local basis $\{\ev_\phi, \ev_\theta\}$, the components of the tangential order-tensor $\Qv_s$ form the matrix
\[
[\Qv_s]  =
\left(
\begin{array}{cc}
q_0 & q_m \\
q_m & -q_0
\end{array}
\right),
\]
where  $q_0(\phi,\theta)$ and $q_m(\phi,\theta)$ are two scalar fields the admissible values of which lie within the disc centered at the origin of the $q_0q_m-$plane with  radius  $1/2$  \cite{Kralj:2011}.   Any in-plane directional ordering is lost wherever  both $q_0$ and $q_m$ vanish. If this is the case, the nematic is in the two-dimensional isotropic phase. Otherwise, the unit eigenvector of $\Qv_s$ corresponding to the greatest eigenvalue yields the average direction of the molecules. Following  \cite{Kralj:2011},   the angle contained between such an average direction and the principal direction $\ev_\phi$ on the sphere is equal to  $  [\arctan(q_m/q_0)]/2$ if $q_0\geq0$, or $  [\pi+\arctan(q_m/q_0)]/2$ if $q_0<0$. 

Earlier studies have shown how the extrinsic curvature of the shell  affects  the ordering of the molecules  \cite{naveprl:2012, nave:2012}. In particular, it has been proven  that  uniform in-plane isotropic states, {\it i.e.} states for which $\Qv_s$ vanishes everywhere, may occur only on surfaces with zero local asphericity or zero mean curvature \cite{NaVe:13}. Uniform flat isotropic states may then occur on the sphere and  minimal surfaces. In this Letter, we study the effects of  lowering  the temperature in a nematics confined to a spherical shell. Initially the temperature is high enough that there is no  in-plane preferred direction.   The nematic order occurs as soon as the temperature reaches a critical value which we shall determine exactly by employing  the surface Landau-de Gennes potential.

This potential is in the form $W = \int_S w \d a$, where $S$ is a sphere of radius $r$ and  the surface energy density $w$ is the sum of  the surface elastic  ($w_{el}$)  and the Landau-de Gennes ($w_{LdG}$) energy densities  \cite{nave:2012}.

Within the one-constant approximation, $w_{el}$ takes the form
\ba
w_{el} = \frac{k}{2}\left( |\grads q_0 + 2 q_m \bom|^2 +  |\grads q_m - 2 q_0 \bom|^2\right) \nonumber \\+  \frac{k}{r^2}(q_0^2 + q_m^2),
\label{eq:wel}
\ea
where $k>0$ is the reduced elastic stiffness,  and $\bom$ is the vector parametrizing the spin connection on the sphere \cite{Giomi:2009}. In spherical coordinates, the vector parametrizing the spin connection and the surface gradient operator  read $\bom = r^{-1}\cot \theta \ev_\phi$  and $\grads=[(r \sin \theta)^{-1}{\pt_\phi}] \ev_\phi+ [r^{-1}{\pt_\theta}]\ev_\theta$, respectively.

 The Landau-de Gennes energy density $w_{LdG}$ is quartic in $q_0$ and $q_m$ and given by
\ba
w_{LdG} = \frac{a}{2} (q_0^2 + q_m^2) + \frac{c}{4} (q_0^2 + q_m^2)^2,
\label{eq:wLdG}
\ea 
where $a=a_0(T-\TNI)/\TNI$, $a_0$ and $c$ are positive constants and $T$ denotes the absolute temperature. The quantity $\TNI$ represents the nematic-isotropic transition temperature for  planar nematics \cite{nave:2012}. 


We now introduce the complex variable $q = q_0 + i q_m$ so that   the total  energy density $w$  becomes
\ba
w = \frac{k}{2}\left|\grads q - 2 i q \bom \right|^2 + 	\underbrace{\frac{a_\mathrm{eff}}{2} |q|^2 + \frac{c}{4} |q|^4}_{\equiv w_{LdG}^{\mathrm{(eff)}}},
\label{eq:energy}
\ea
where ${a}_\mathrm{eff} = a + 2 k/r^2$ depends not only on the constitutive properties of the nematics, but also on the extrinsic curvature of the sphere. Thus, the energy density  $w_{LdG}^{\mathrm{(eff)}}$ can  be regarded  as the effective thermal potential on the sphere. 

Following the variational scheme introduced in \cite{nave:2010}, the (dimensionless) Euler-Lagrange equation associated with the surface Landau-de Gennes potential is found to be 
\ba
\mathcal{L} q   - \ab q - \cb |q|^2 q=0,
\label{eq:EL}
\ea
where $\mathcal{L}=r^2(\Delta_s  - 4 i \bom \cdot \grads  - 4  |\bom|^2)$,  $\Delta_s=(r\sin\theta)^{-2}\partial^2_{\phi}+r^{-2}(\sin\theta)^{-1}\partial_\theta(\sin\theta\partial_\theta)$ is the Laplace-Beltrami operator on the sphere, $\ab=a_\mathrm{eff}/k$ and $\cb=c/k$.
It is immediate to check that  \eqref{eq:EL}  admits the trivial solution $q=0$ (which corresponds to the in-plane isotropic state) for any values of $\ab$ and $\cb$. However, as the temperature decreases, $\ab$ becomes smaller and smaller, and, as soon as it reaches a critical value, the trivial solution becomes unstable. The onset of  nematic ordering  occurs then as a bifurcation from the trivial solution. 

To study this bifurcation problem, we expand $q$ as a  power series in the dimensionless parameter $\eps = \sqrt{\ab/\abc-1}\ll1$ that  provides a measure  of the small departures of $\ab$ from its value $\abc$ at the nematic-isotropic transition temperature $\Tc$ on the sphere. For small temperature departures  from $\Tc$, we then  look for solutions of \eqref{eq:EL} in the form  $q= \sum_{n=1}^{+\infty}\eps^n q_n $.   

To leading order, $q_1$ satisfies the linear equation  
\ba
\elle q_1 =\ab q_1.
\label{eq:SL}
\ea
Since the differential operator $-\mathcal{L}$ is self-adjoint and positive,  its spectrum is countable and its eigenvalues are real and positive. More precisely, the spectrum of $-\elle$ is the set $\{\lambda_n=n^2+n-4:n\in\mathbb{N},n\geq 2\}$, whence  $\lambda_2=2$ is the least eigenvalue of $-\elle$. As a direct consequence,  equation \eqref{eq:SL} admits a non-trivial solution if and only if $\ab\leq\abc=-2$, which implies that  the in-plane isotropic state $q=0$ is   stable if and only if $\ab>\abc=-2$ or, equivalently in view of the definitions of $a$ and $a_\mathrm{eff}$,
\begin{equation}\label{tni}
T>\Tc=\TNI\left(1-\frac{4k}{a_0r^2}\right).
\end{equation}
The extrinsic curvature curvature has then the effect to make  the nematic-isotropic transition temperature on the sphere be lower than that for planar nematics.

The smooth eigenfuntions of $-\elle$ corresponding to the least eigenvalue $\lambda_2=2$  can be written in terms of complex  trigonometric functions (normalized on using  the Hermitian inner product $\left\langle u,v\right\rangle =\int_\esse uw^\star \d a$, where the superscript $^\star$ denotes the complex conjugate) as  
\begin{equation}\label{eigfun}
\left.\begin{array}{ll}
f_{-2} = \displaystyle\frac12 \sqrt{\frac{5}{\pi}} \exp({-2 i \phi}) \sin^4 \frac\theta2,\\
[3mm]
 f_{-1} = \displaystyle\sqrt{\frac{5}{\pi}} \exp({- i \phi}) \sin^3 \frac\theta2 \cos\frac\theta2,\\
[3mm]
 f_0 = \displaystyle\sqrt{\frac{15}{2\pi}} \sin^2 \frac\theta2\cos^2\frac\theta2,\\
[3mm]
 f_{1} = \displaystyle \sqrt{\frac{5}{\pi}} \exp({i \phi}) \cos^3 \frac\theta2 \sin\frac\theta2 ,\\
[3mm]
 f_{2} =  \displaystyle\frac12\sqrt{\frac{5}{\pi}} \exp({2 i \phi}) \cos^4 \frac\theta2. 
\end{array}\right.
\end{equation}

The bifurcate solutions are then  linear combinations of these five modes
\ba
q_1= \sum_{h=-2}^2 A_{h} f_{h},
\label{eq:q1}
\ea
where $A_h=\varrho_h\exp(i \psi_h)$ are arbitrary complex constants.  These  are to be determined by taking into account  the non-linear term in \eqref{eq:EL} that has thus far been  neglected. Taking into account that $\abc=-2$ and  the definition of $\eps$,  for small temperature departures  from the critical value \eqref{tni}  we have $\ab=-2(1+\eps^2)$. Thus, collecting terms of order $O(\eps^3)$ in the equilibrium equation \eqref{eq:EL}, with  $q$ expanded   as a power series in  $\eps$, gives the  equation $(\elle+2)q_3=(2-\cb|q_1|^2)q_1\equiv g$. On using  the Lyapunov-Schmidt reduction \cite{rand} and in view of the self-adjointness of $\elle$, this equation is solvable providing that  $g$ belongs to the orthogonal complement of the kernel of $\elle +2$. In other words, we have to require that $\left\langle g,f_h\right\rangle=0$ for all $h=-2,...,2$. These conditions  yield a system of five nonlinear algebraic equations in the five unknowns $A_h=\varrho_h\exp(i\psi_h)$.  

On the other hand, the Poincar\'e-Hopf index theorem guarantees that on a spherical nematic shell there must be at least one melting point. Therefore,  without loss of generality, we can limit our bifurcation  analysis to linear combinations \eqref{eq:q1}  that vanish  at the north pole, i.e.  for $\theta=0$. This ansatz reduces by one the number of unknowns and equations in the system of nonlinear algebraic equations $\{\left\langle g,f_h\right\rangle=0:h=-2,..,2\}$. Indeed, since  $f_2$ does not vanish at the north pole we  take $A_2=0$ in \eqref{eq:q1} and release the condition $\left\langle g,f_2\right\rangle=0$.  The resulting reduced system of  nonlinear algebraic equations admits infinitely many solutions, each of which determines uniquely a bifurcate texture.  

Within the  Landau-de Gennes theory, defects are melting points, i.e. isolated points where $q$ vanishes. According to the number of the melting points and their distribution on the spherical shell, the bifurcate textures can be divided into seven broad classes of configurations.
\begin{itemize}
 \item [($\mathcal{C}_1$)] Monovalent: a single $m=+2$ melting point. 
 \item [($\mathcal{C}_2$)] Linear non-axisymmetric: two antipodal melting points  with topological charges $+1/2$ and  $+3/2$.  
 \item [($\mathcal{C}_3$)] Linear axisymmetric: two $m=+1$ antipodal melting points. 
\item [($\mathcal{C}_4$)] Trigonal: three melting points located at the vertices of an isosceles triangle on a maximum circle on the sphere, two of them with $m= +1/2$ and the other with $m=+1$. The base angles of the  isosceles triangle measure $\alpha_1=\mathrm{arctan}\sqrt[4]{40}\approx 68.315^\circ$ and the vertex angle measures $\alpha_2=\pi-2\mathrm{arctan}\sqrt[4]{40}\approx 43.37^\circ$. The values of  $\alpha_1$ and $\alpha_2$ are in perfect agreement with the experimental results reported in \cite{Lopez-Leon:2011}.
  \item [($\mathcal{C}_5$)] Tetrahedral:  four $m=+1/2$ melting points located at the vertices of a regular tetrahedron.
 \item [($\mathcal{C}_6$)] Squared: four $m=+1/2$ melting points  situated at the vertices of a square on a maximum circle of the sphere. 
 \item [($\mathcal{C}_7$)] Rectangular: four $m=+1/2$ melting points  sitting at the vertices of a rectangle on a maximum circle on the sphere. The bifurcate textures belonging to this class are defined by a scalar parameter $\varrho_0$ (see Table \ref{tab:table1}). The two limiting configurations $\varrho_0\rightarrow0$ and $\varrho_0 \rightarrow \sqrt{(28\pi)/(5\bar{c})} $  correspond to $\mathcal{C}_6$ and  $\mathcal{C}_3$, respectively. Thus, as $\varrho_0$ tends to zero  the melting points tend to locate at the vertices of  a square on a maximum circle on the sphere, while   as $\varrho_0$ increases the four $+1/2$  defects move on a maximum circle, approaching each other two by two until they collapse into two $+1$ defects. 
\end{itemize}

As highlighted  in Table \ref{tab:table1}, these classes of bifurcate configurations are linear combinations of one, two or at most three of the normal modes \eqref{eigfun}. All the bifurcate configurations but $\mathcal{C}_7$ are    characterized solely by the moduli $\varrho_h$ of the  complex coefficients $A_h$, while the    arguments $\psi_h$ are  arbitrary.  For the rectangular  configuration (that is the only bifurcate state resulting from the linear combination of  three normal modes; specifically, $f_0$, $f_1$ and $f_{-1}$), the moduli of the coefficients $A_0$, $A_{1}$ and $A_{-1}$ are as in Table \ref{tab:table1}, while the arguments  are subject to the restriction  $2\psi_0-\psi_1-\psi_{-1}=(2l+1)\pi$, with $l\in\mathbb{Z}$. Therefore, for rectangular configurations only  two out of the three arguments  are completely arbitrary.  

For the linear axisymmetric configuration the nematic texture varies as the argument $\psi_0$ varies. Indeed, $\psi_0=0$ corresponds to a uniform alignment along the parallels (purely bend phase), while $\psi_0={\pi}/{2}$ corresponds instead to a uniform alignment along the meridians (purely splay phase).   For the monovalent and linear non-axisymmetric configurations the  arbitrariness  of the arguments  reflects    the invariance of these states   under rigid rotations  around the polar axis.   Finally, for the  bifurcate configurations resulting from a linear superposition of two or three normal modes,  it can be proven that the invariance   under rigid rotations  around the polar axis reflects on the arbitrariness of one of the two arbitrary arguments, the other determines instead the texture.

 As the determination of the topological charge $m$ of a melting point located at the point $p$ is concerned, we  follow similar arguments as in \cite{Rosso:2012} and find out that $m=m_\varphi+\oint_{\mathscr{C}} [\arg(q)]' \d s\big/(4\pi)$, where $\mathscr{C}$ is a regular closed simple circuit that is oriented anti-clockwisely around the normal to the sphere at $p$ and can be continuously contracted to $p$, $s$ denotes the
arc-length  along ${\mathscr{C}}$, and the prime denotes differentiation  with respect to $s$. The term $m_\varphi$ is instead the topological charge of the reference field $\ev_\varphi$, which vanishes wherever $\ev_\varphi$ has no singularity, that is at any point on the sphere except for the poles where it is equal to unity.

\begin{table*}
  \centering
    \begin{tabular}{|c|c|c|c||c||c|} 
		\hline
		\multicolumn{4}{|c||}{Moduli of $A_h$}    & Melting points & Dimensionless energy\\
		\hline
      $\varrho_{-2}$ & $\varrho_{-1}$ & $\varrho_0$ & $\varrho_1$  & {Topological charge and location} & $W\bar{c}/(k\pi)$\\
      \hline
      \hline
      %
      %
      \footnotesize{$ \displaystyle{\frac{6}{5}\sqrt{\frac{2\pi}{\bar{c}}}}$} & - & - & -     & \raisebox{-1.1cm}{\includegraphics[width=2.0cm,keepaspectratio]{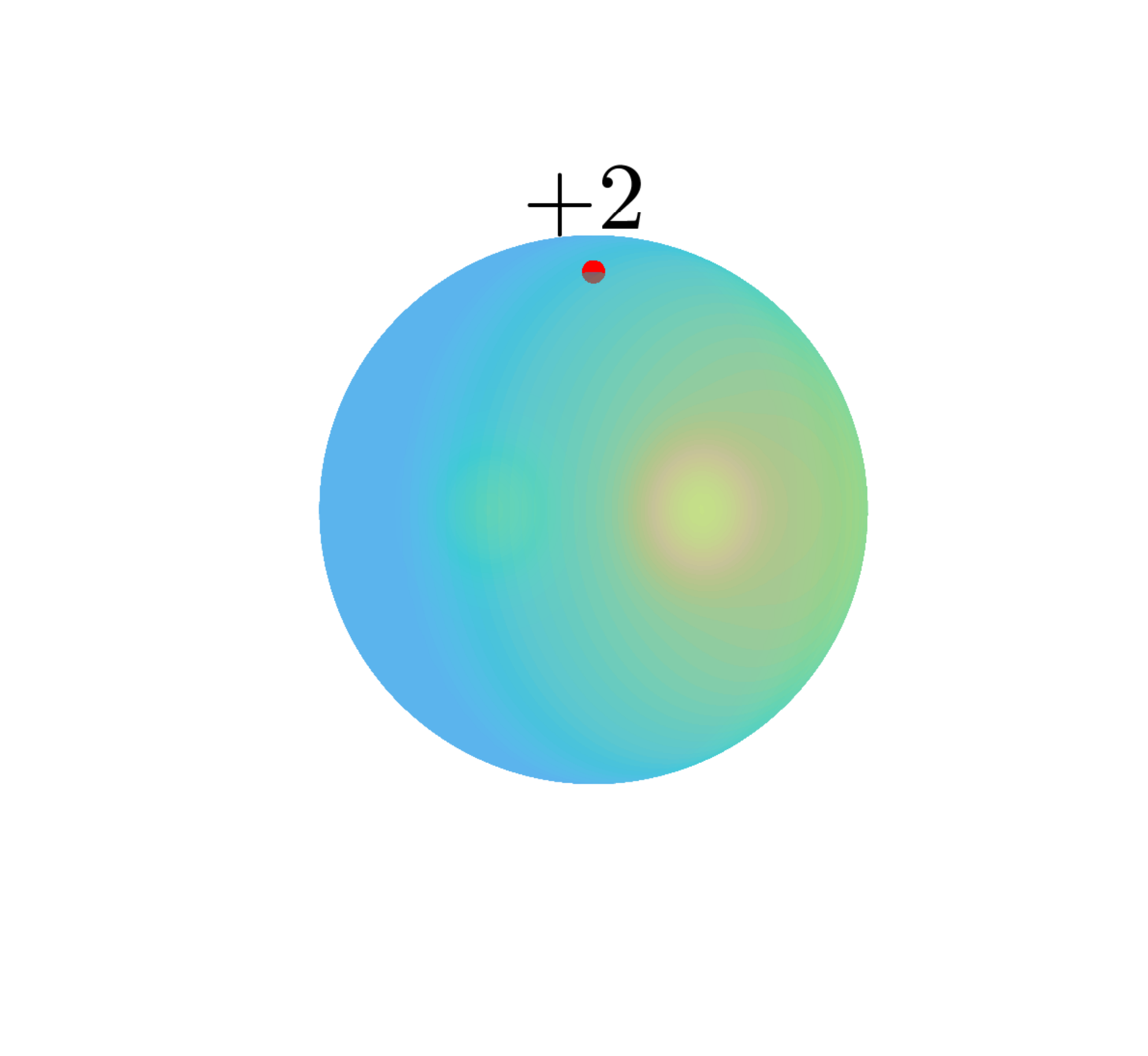}} & $-\displaystyle \frac{36}{25}\eps^4$ \\
       \hline
      - & \footnotesize{$\displaystyle\frac35\sqrt\frac{14\pi}{\bar{c}}$ } & - & -  &  \raisebox{-1.1cm}{\includegraphics[width=2.0cm,keepaspectratio]{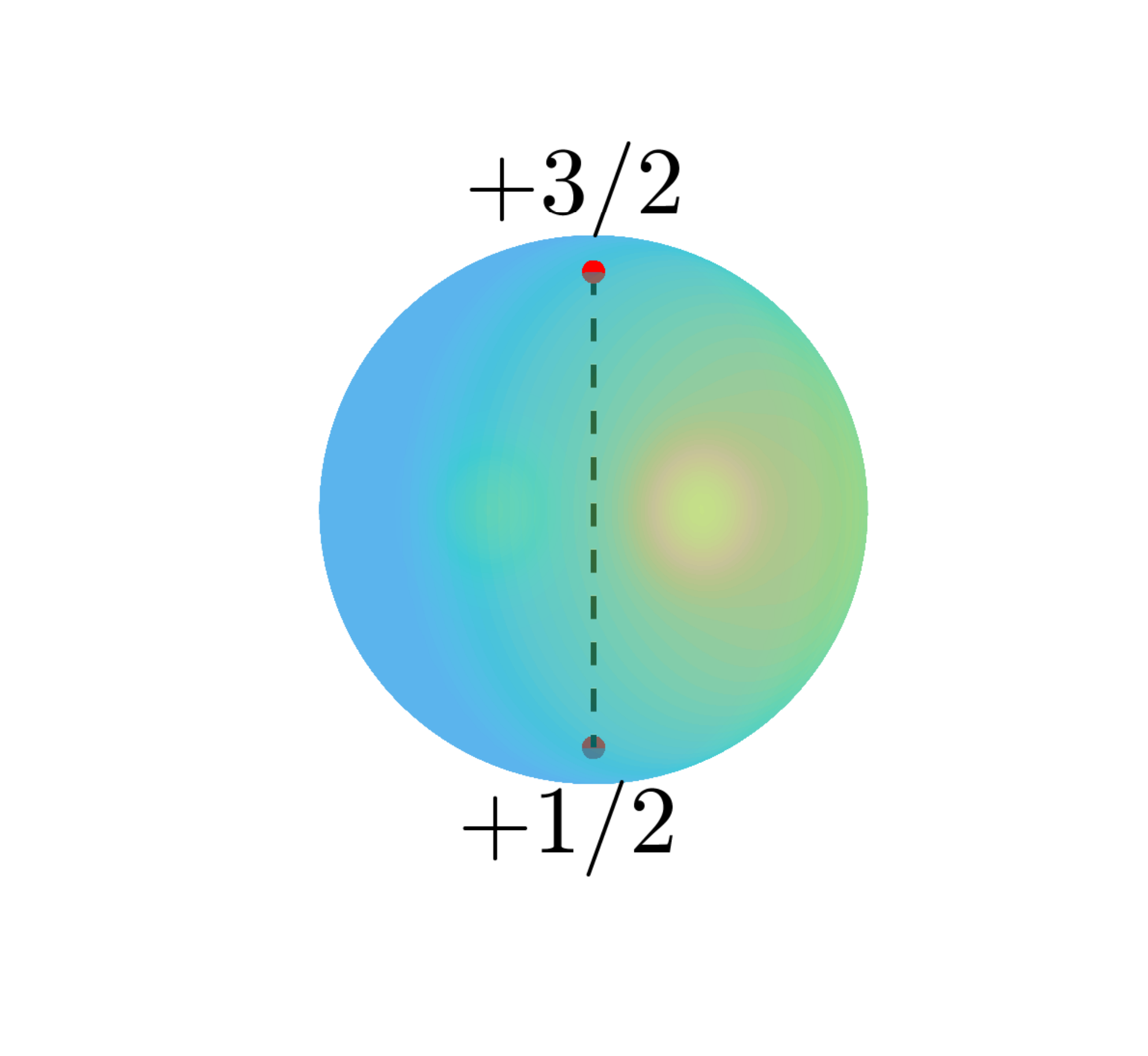}} & $-\displaystyle \frac{63}{25}\eps^4$\\
       \hline
      - & - & \footnotesize{$\displaystyle\sqrt\frac{28\pi}{5\bar{c}}$}  & -   & \raisebox{-1.1cm}{\includegraphics[width=2.0cm,keepaspectratio]{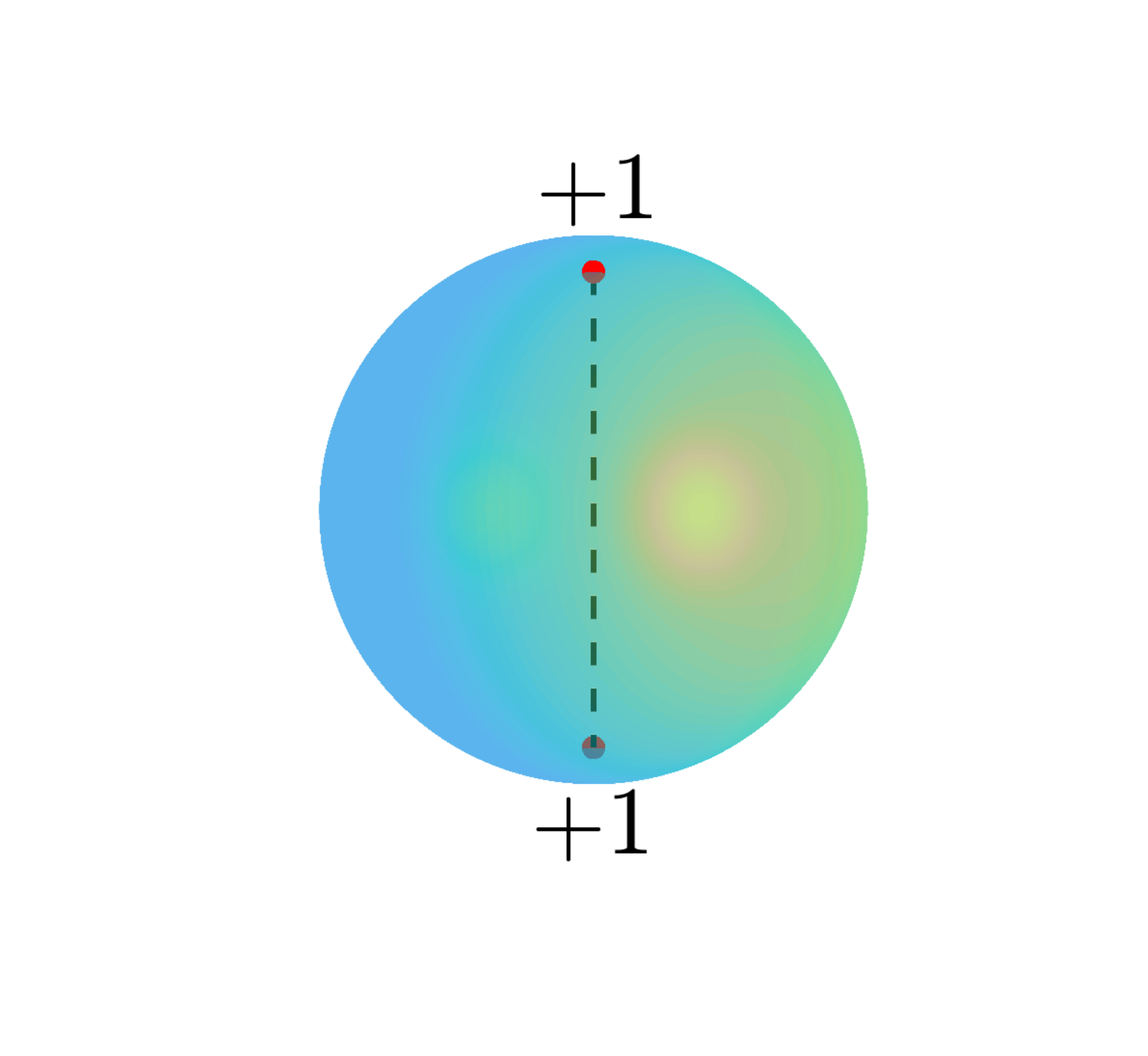}} & $-\displaystyle\frac{14}{5}\eps^4$ \\
       \hline
       - & - & - &  \footnotesize{$ \displaystyle\frac35\sqrt\frac{14\pi}{\bar{c}}$}   & \raisebox{-1.1cm}{\includegraphics[width=2.0cm,keepaspectratio]{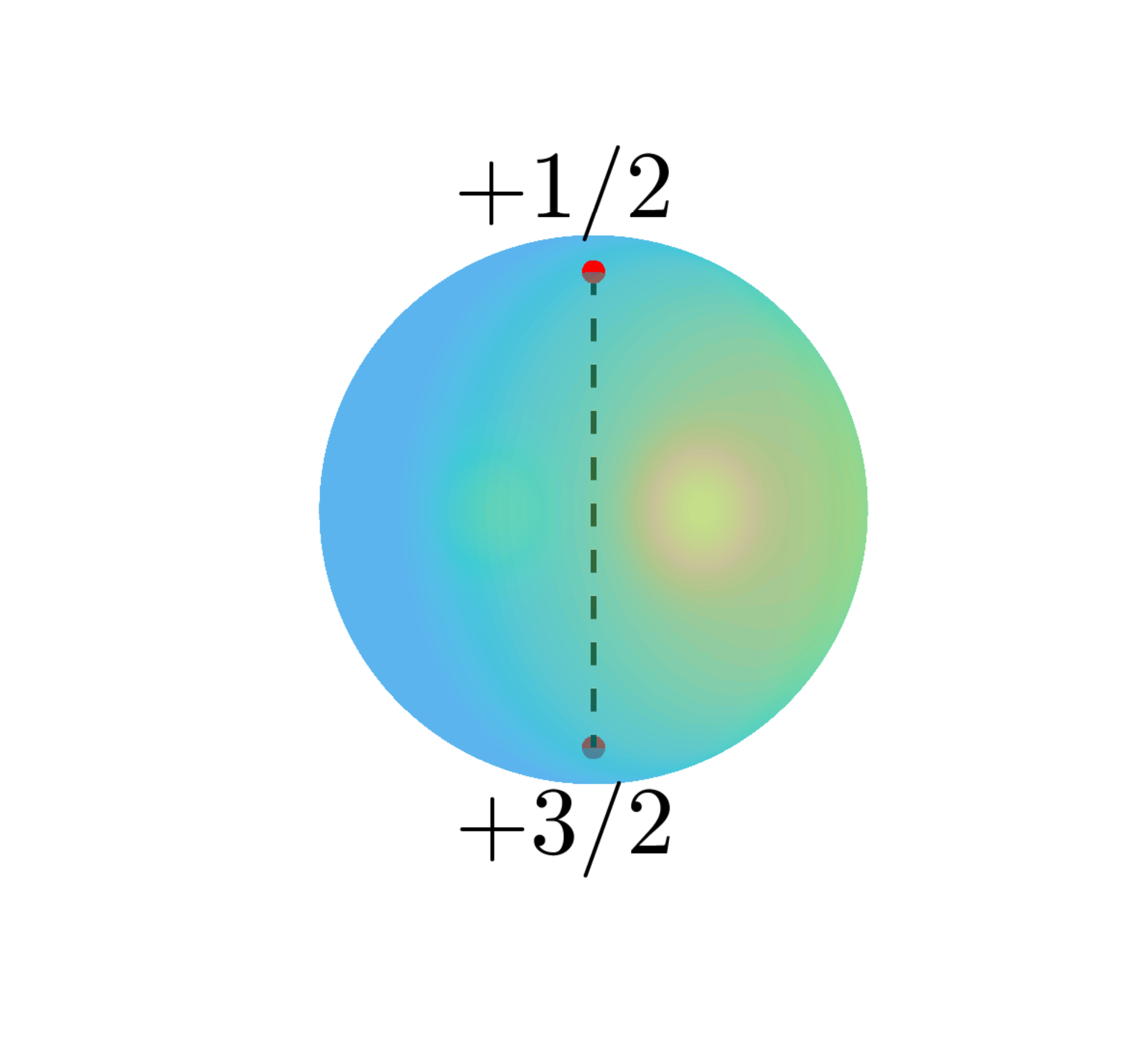}} & $-\displaystyle \frac{63}{25}\eps^4$ \\
      \hline
       \hline
      %
      %
      %
			\footnotesize{$\displaystyle\frac{2}{5}\sqrt{\frac{14 \pi}{3\bar{c}}}$} &- &\footnotesize{$\displaystyle\frac{4}{3}\sqrt{\frac{14 \pi}{5\bar{c}}}$} & - & \raisebox{-0.8cm}{\includegraphics[width=1.6cm,keepaspectratio]{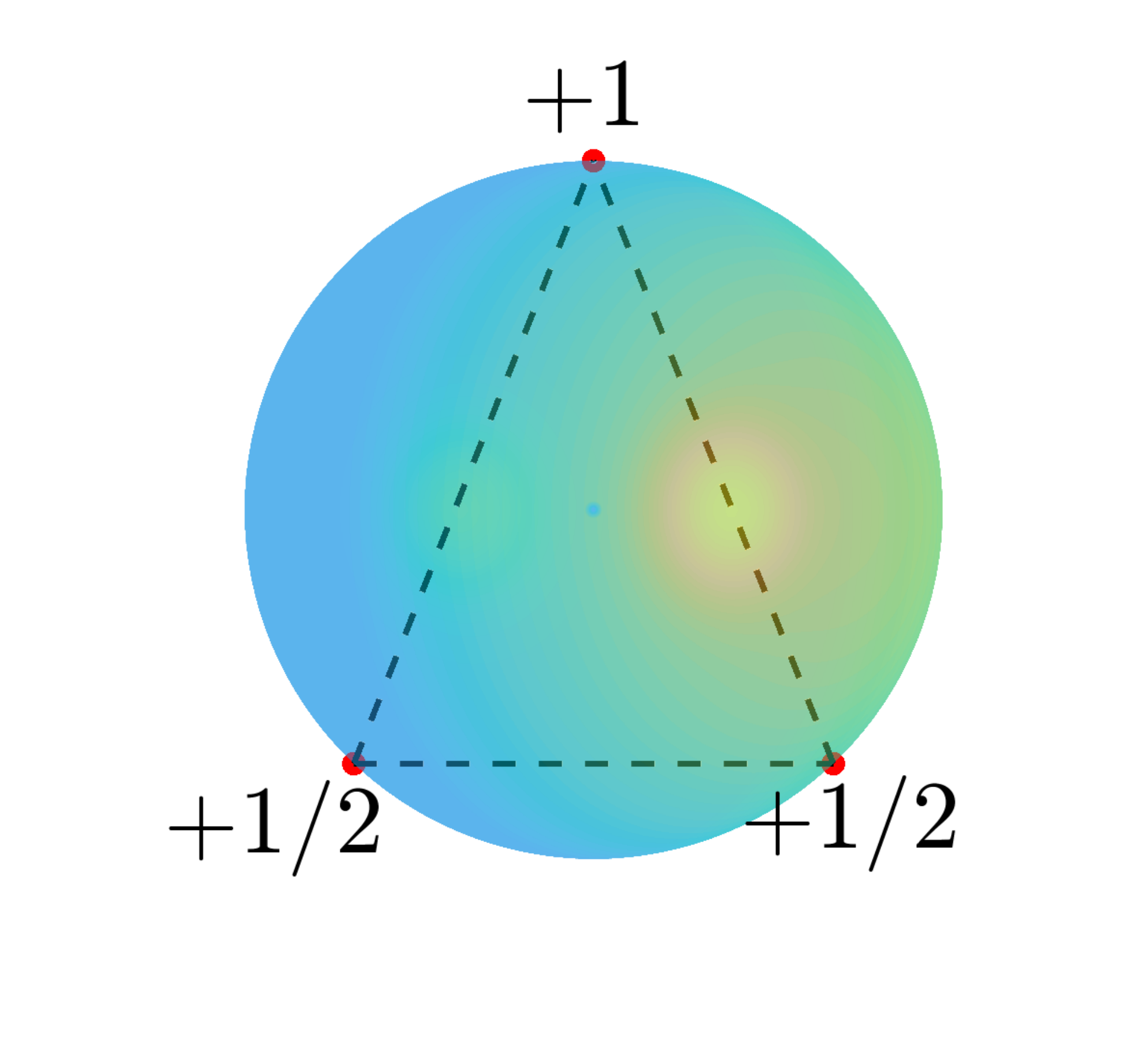}} & $-\displaystyle \frac{644}{225}\eps^4$\\
      \hline
       \footnotesize{$\displaystyle\frac{2}{5} \sqrt{\frac{14\pi}{\bar{c}}}$} & - & - &  \footnotesize{$\displaystyle\frac{4}{5}\sqrt{\frac{7\pi}{\bar{c}}}$}   & \raisebox{-1.1cm}{\includegraphics[width=2.0cm,keepaspectratio]{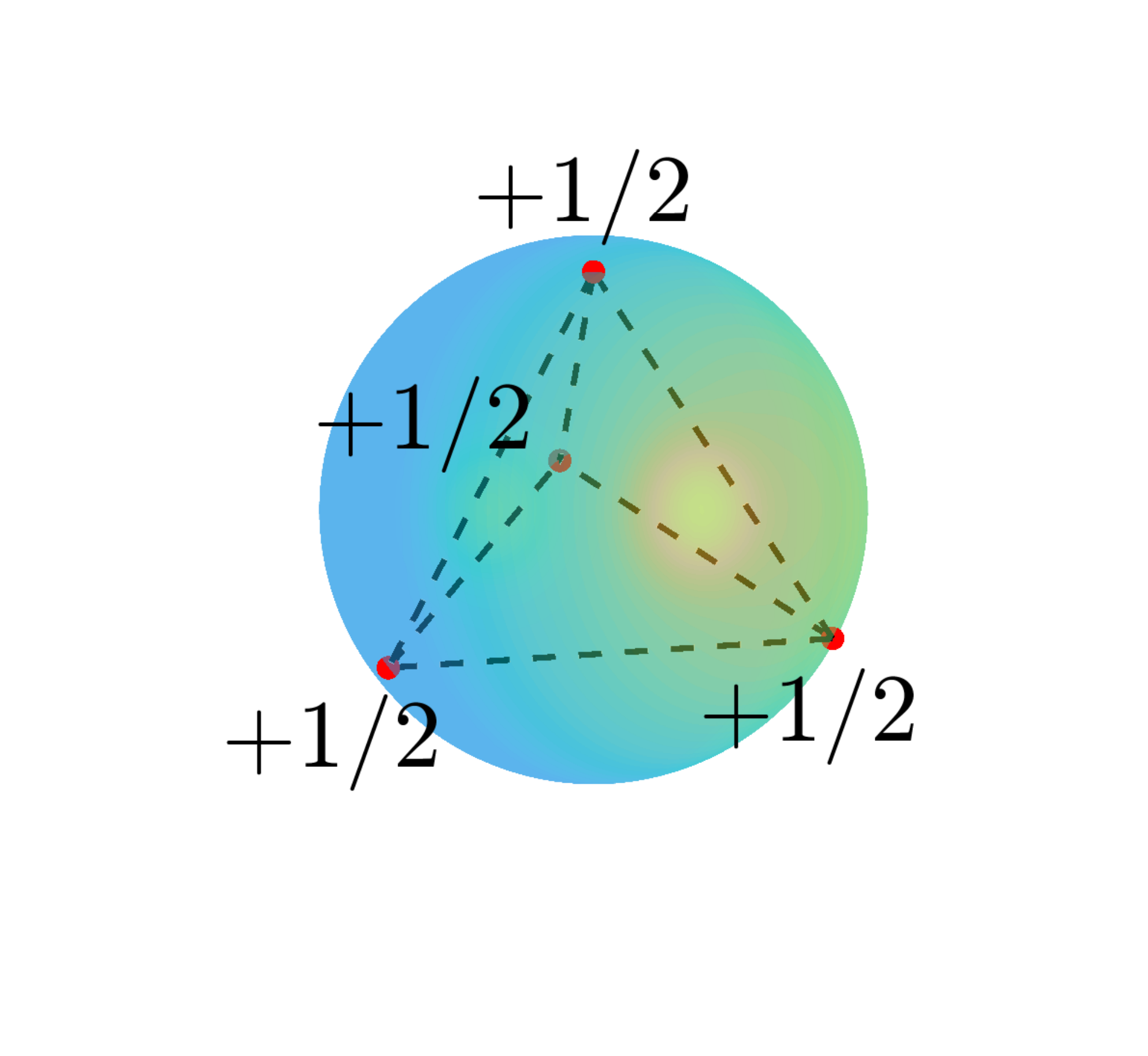}} & $-\displaystyle \frac{84}{25}\eps^4$ \\
      \hline
      - &$ \displaystyle\sqrt\frac{14\pi}{5\bar{c}}$   & - &$ \displaystyle\sqrt\frac{14\pi}{5\bar{c}}$       & \raisebox{-0.8cm}{\includegraphics[width=1.6cm,keepaspectratio]{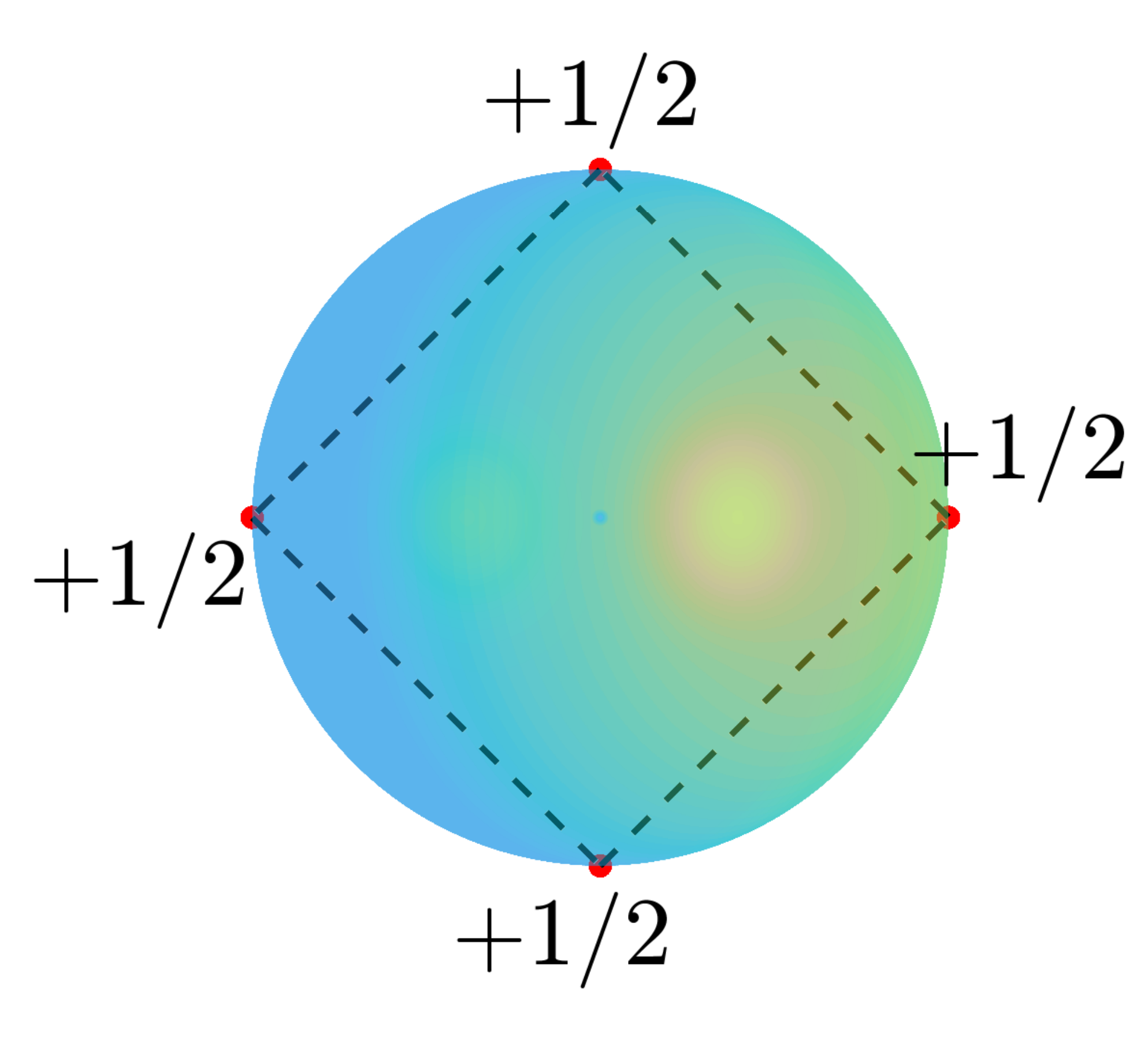}} & $-\displaystyle\frac{14}{5}\eps^4$\\
			\hline
      \hline
      %
        - &\footnotesize{ $\displaystyle\sqrt\frac{28\pi - 5 \varrho_0^2 \bar{c}}{10 \bar{c}}$ }&\footnotesize{$ \in\left]0,\displaystyle\sqrt\frac{28\pi}{5\bar{c}}\right[$ }& \footnotesize{$\displaystyle\sqrt\frac{28\pi - 5 \varrho_0^2 \bar{c}}{10 \bar{c}}$} &  \raisebox{-0.8cm}{\includegraphics[width=1.6cm,keepaspectratio]{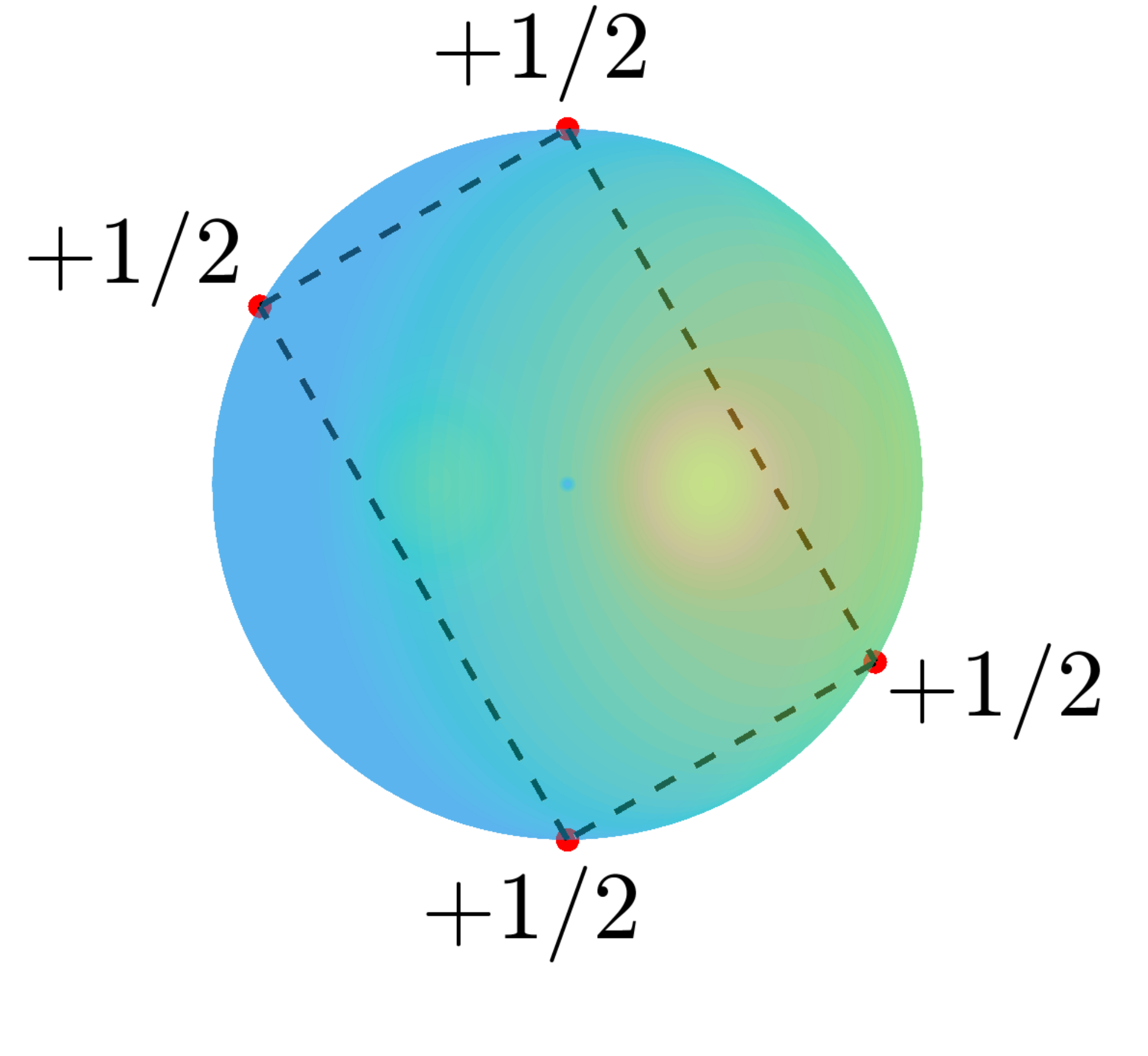}} & $-\displaystyle\frac{14}{5}\eps^4$\\
      \hline
       \end{tabular}
 \caption{The table is divided into three horizontal blocks (separated by  double lines) relating, from top to bottom, to solutions resulting from the linear superposition of one, two and three normal modes. The first  four columns report the values of the moduli $\varrho_h$ of the complex coefficients in the linear combinations. The second last column displays  locations of melting points on the spherical shell; the last column contains the dimensionless energies of the bifurcate configurations. \label{tab:table1}}
\end{table*}

\begin{figure}
	\centering
		\includegraphics[width=7cm,keepaspectratio]{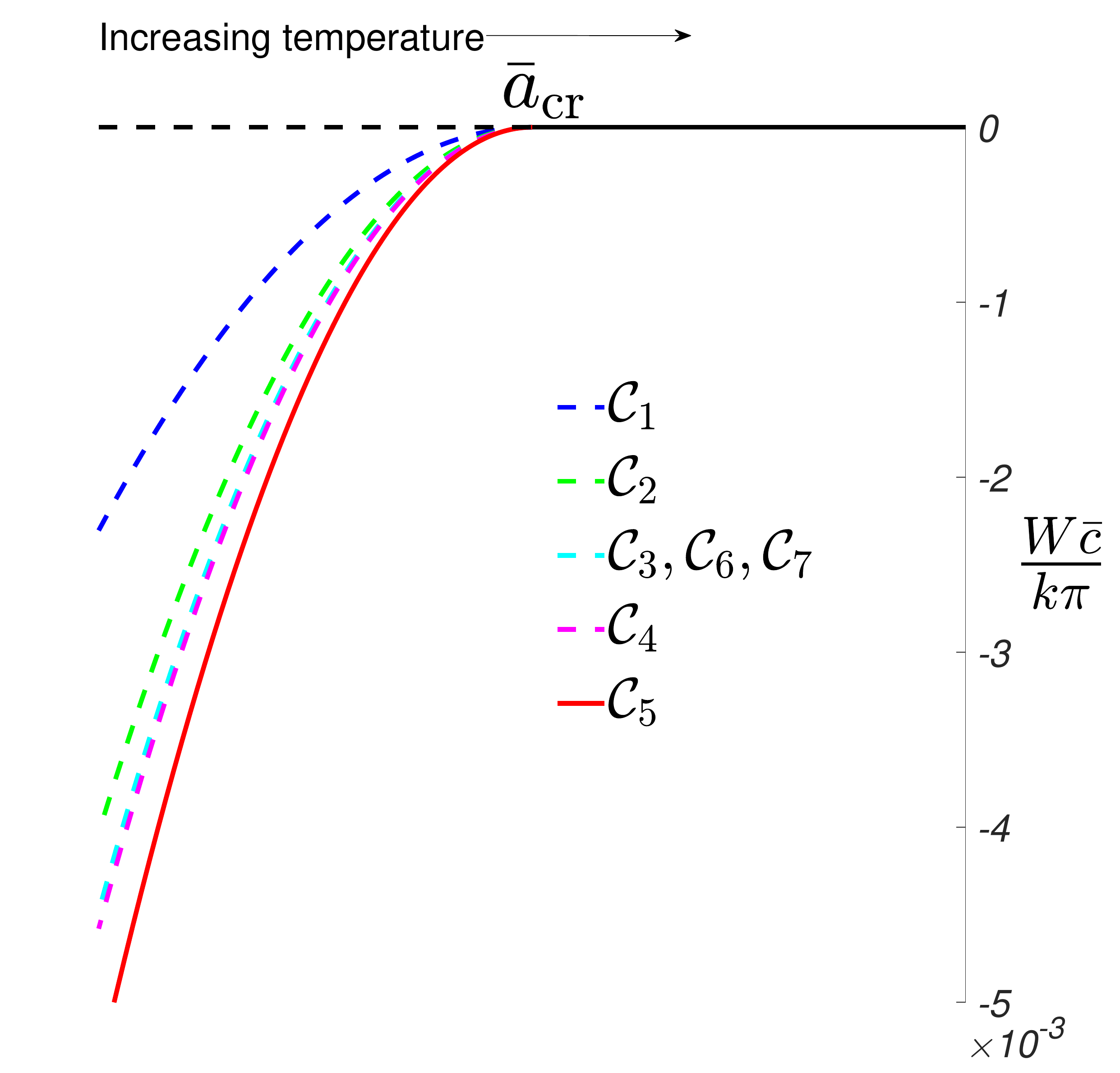}
	\caption{Energies of the bifurcate configurations. For small temperature  departures from the critical value $T_\mathrm{cr}$, the tetrahedral configuration ($\mathcal{C}_5$) is  a local minimizer of the energy functional. All the other configurations are  unstable. \label{fig:energies}}
\end{figure}

The  dimensionless energies of  the bifurcate configurations are reported in Table \ref{tab:table1}. In agreement with the theoretical result  in \cite{Lubenski92}, the tetrahedral configuration $\mathcal{C}_5$ is the ground state. It is worth noting that the rectangular, squared and linear axisymmetric  states have the same energy (see Figure \ref{fig:energies}). This implies  that, at small temperature departures from $\Tc$, the rectangular configuration can vary continuously from the squared configuration to the linear axisymmetric one with no energy cost.

To determine   the locally stable bifurcate solutions, we study the energy functional at temperatures slightly lower than  $\Tc$. As observed above, at these temperatures the approximate equilibrium solutions are of the form \eqref{eq:q1} with $A_2=0$. This  enables us to adopt the analytical approach introduced in \cite{Barbero:2000}. Upon substitution of \eqref{eq:q1} and \eqref{eigfun} into the energy functional, the stability analysis reduces to the determination of the local minima of a real-valued function $\mathcal{W}$ in the four complex variables  $A_{-2}$ ,$A_{-1}$, $A_0$ and $A_1$. The stationary points of $\mathcal{W}$ lead to  the same equilibrium configurations that  we  have determined by means of the Lyapunov-Schmidt reduction. Therefore, the study of the positive definiteness of the Hessian matrices  at the bifurcate configurations can tell us which bifurcate configuration is a local minimizer of $\mathcal{W}$. We have found  that only the tetrahedral configuration is a local minimizer of $\mathcal{W}$. This means that only $\mathcal{C}_5$ is stable, the others are unstable (Figure \ref{fig:energies}). 

 The instability of the monovalent and linear non-axisymmetric configurations is not surprising as they have never been observed experimentally in non-chiral nematic shells, whereas they have been theoretically predicted and experimentally observed in chiral nematic shells \cite{Darmon:2016, Darmon:2016SM}. The linear axisymmetric configuration is unstable for non-axisymmetric perturbations which lead it towards the tetrahedral state. This  is in agreement with the results in \cite{Vitelli:2006}. It must be said that the bivalent axysimmetric configuration can be stabilized  by the non-uniformity of the shell thickness \cite{Koning:2013}, or out-of-plane escapes of the director \cite{Vitelli:2006,napoli:2021}.  Both aspects have  not been considered in this paper. When the nematic director field is not constrained to be tangential, the configuration with two $+1$  disclinations  is the ground state for sufficiently small particles \cite{napoli:2021}.

In summary, we use the two-dimensional Landau-de Gennes theory to study the onset of the temperature-induced nematic ordering in spherical shells which are initially in the isotropic phase. This approach has the advantage of predicting the transition temperature as well as  determining exactly the textures just below the critical temperature.

  The critical temperature depends on the size of the particle (c.f. equation \eqref{tni}). Indeed, the extrinsic curvature  lowers the isotropic-nematic transition temperature by a quantity proportional to the  square of the ratio between the nematic coherence length $\sqrt{k/a_0}$ and the shell radius $r$.  This effect is however negligible in the shells that are currently produced and used in the experiments. Typical  radii of the nematic shells are of order of tens to hundreds of micrometers \cite{Lopez-Leon:2011, Noh:2020}, while the nematic coherence length is of order  of nanometres. To appreciate the cooling effect of the extrinsic curvature,  the shell radius should be at least one order of magnitude larger than the nematic coherence length. If this were the case, the extrinsic curvature of the shell would lower the isotropic-nematic transition temperature by 1\%.

Far more interesting are the results  on the nematic alignment at the critical temperature. The textures  we have obtained by solving a generalized Thomson-like problem are expressed in closed form in terms of elementary trigonometric functions and  reproduce different types of configurations. Some of these have been already predicted theoretically,  found numerically, or observed experimentally. The total topological charge of the melting points obeys the Poincar\'e-Hopf theorem. The energy and stability analysis confirms that the tetrahedral configuration represents the ground state. Finally, the exact solutions we have here obtained can definitely be used as initial guesses in numerical investigations of more complex theories in which the fusion of the defect core and out-of-plane escape is considered simultaneously \cite{Susser:2020, napoli:2021}.


\subsection*{Acknowledgements}
 This research has been partially supported by  GNFM of Italian INDAM and by the PRIN 2017 research project  (n. 2017KL4EF3) ``Mathematics of active materials: from mechanobiology to smart devices''.

\end{document}